\documentclass[11pt, a4paper, logo, copyright]{googledeepmind}

\pdftrailerid{redacted}

\makeatletter
\renewcommand\bibentry[1]{\nocite{#1}{\frenchspacing\@nameuse{BR@r@#1\@extra@b@citeb}}}
\makeatother

\usepackage{listings}
\usepackage{kantlipsum, lipsum}
\usepackage{dsfont}
\usepackage{gdm-colors}
\usepackage{float}
\usepackage{xspace}
\usepackage{todonotes}
\usepackage{wrapfig}
\usepackage{cleveref}
\usepackage{adjustbox} 

\usepackage{makecell}
\usepackage{tabularx}
    \newcolumntype{L}{>{\raggedright\arraybackslash}X}

\newcommand{\fullname}[0]{\textbf{Trusted Capable Model Environments}\xspace}
\newcommand{\tmle}[0]{\textbf{TCME}\xspace}
\newcommand{\tmles}[0]{\textbf{TCMEs}\xspace}

\usepackage[many]{tcolorbox}
\usepackage{xspace}
\usepackage{todonotes}

\usepackage{stackrel}

\NewTColorBox{NewBox}{ s O{h} }{%
  floatplacement={#2},
  IfBooleanTF={#1}{float*,width=\textwidth}{float},
  colback=green!5!white,colframe=green!70!black,%
}

\NewTColorBox{NewBoxRed}{ s O{h} }{%
  floatplacement={#2},
  IfBooleanTF={#1}{float*,width=\textwidth}{float},
  colback=red!5!white,colframe=red!70!black,%
}

\usepackage[authoryear, sort&compress, round]{natbib}

\graphicspath{{figures/}}

\title{Trusted Machine Learning Models Unlock Private Inference for Problems Currently Infeasible with Cryptography}

\correspondingauthor{iliashumailov@google.com}

\renewcommand{\today}{}

\author[1]{Ilia Shumailov}
\author[2]{Daniel Ramage}
\author[3]{Sarah Meiklejohn}
\author[2]{Peter Kairouz}
\author[1]{\\Florian Hartmann}
\author[1]{Borja Balle}
\author[2]{Eugene Bagdasarian}

\affil[*]{Inverse alphabetic order}
\affil[1]{Google DeepMind}
\affil[2]{Google Research}
\affil[3]{Google}

\begin{abstract}

We often interact with untrusted parties. Prioritization of privacy can limit the effectiveness of these interactions, as achieving certain goals necessitates sharing private data. 
Traditionally, addressing this challenge has involved either seeking \textit{trusted intermediaries} or constructing \textit{cryptographic protocols} that restrict how much data is revealed, such as multi-party computations or zero-knowledge proofs. While significant advances have been made in scaling cryptographic approaches, they remain limited in terms of the size and complexity of applications they can be used for. 
In this paper, we argue that  %
capable machine learning models can fulfill the role of a trusted third party,
thus enabling secure computations for applications that were previously infeasible. 
In particular, we describe \textbf{\fullname (\tmles)} as an alternative approach for scaling secure computation, where capable machine learning model(s) interact under input/output constraints, with explicit information flow control and explicit statelessness. This approach aims to achieve a balance between privacy and computational efficiency, enabling private inference where classical cryptographic solutions are currently infeasible. We describe a number of use cases that are enabled by \tmle{}, and show that even some simple classic cryptographic problems can already be solved with \tmle{}. Finally, we outline current limitations and discuss the path forward in  implementing them.
\end{abstract}

\begin{document}

\maketitle

\section{What are~\tmles?}

In this paper we contend that recent advancements in machine learning enable a new paradigm for private inference. Fundamentally, the need for many cryptographic primitives stems from the fact that we don't have trusted third parties, thus requiring mutually untrusted participants to interact in a way that avoids revealing their data to each other but where they can nevertheless agree on a result.  In this paper we argue that a capable machine learning model, in some settings, can play the role of a trusted third party~\citep{abadi2004trustedcomputing,anderson2010security}. We propose \textbf{\fullname{}}, a setting where an individual machine learning model or a number of models initiate an interaction with additional constraints in the input and output to ensure that private data cannot leave the \tmle. Consider for example the classical millionaires problem, where a pair of individuals are trying to figure out who has more money without disclosing how much money they have. Cryptographically, this can be solved using a secure two-party computation by Yao, or subsequent protocols that are more efficient~\citep{yao1982protocols,ioannidisgrama2003efficientyao,lin2005efficient}. With \tmle{}, both individuals agree on 
\begin{itemize}
    \item \textbf{A model}, e.g., \texttt{Gemma}~\citep{gemmateam2024gemmaopenmodelsbased};
    \item \textbf{A prompt}, e.g., ``Say "first" if A is bigger than B and "second" otherwise? A=\{\} and B=\{\}.'';
    \item \textbf{Input constraints}, e.g., A and B are 32-bit integers;
    \item \textbf{Output constraints}, e.g., the only approved outputs are `first' or `second'.
\end{itemize}
If the environment can be trusted to both avoid leaking the private information provided by each party and to reliably output the correct answer, this presents an alternate approach to enabling this computation. While a viable cryptographic solution exists for this simple example, for the applications we discuss below it is currently computationally infeasible to rely on cryptographic solutions due to the unstructured nature of the computation. In contrast, we argue that securely performing these computations is entirely feasible with a new inference paradigm utilising machine learning models. 

\emph{\textbf{This paradigm for private computations enables analysis and collaboration for tasks that were previously infeasible.}} For example, programming the task is no longer limited to a highly technical specification, where all possible states have to be modeled; instead, it is possible to use human language directly by non-specialists. 

There are three fundamental properties that we need to satisfy in order for models to be trusted:
\begin{enumerate}
    \item \textbf{Statelessness} -- the model should be incapable of memorising, learning, or retaining any state based on the data with which it interacts. That way, it is clear with each interaction what (private) data influences the output, ensuring confidence in that post-invocation no private data can leak from the model and the model can't discriminate the user based on prior interactions. 
    \item \textbf{Explicit Information Flow Control}
    \begin{itemize}
    \item \textbf{Information Flow} -- the model and the underlying system should have an explicit and frozen information flow that can be defined in a coordinated (and verifiable) way. 
    \item \textbf{Verifiability} -- users of the~\tmle{} need an explicit mechanism for verifying that the correct model, correct prompt, and input/output constraints are respected. 
    \end{itemize}
    \item \textbf{Trustworthy and Capable Model(s)} -- we assume the use of \emph{trustworthy} model(s) that are \emph{capable} of solving a given user task, and that the model(s) are aligned in their performance with the expectations of the involved parties. 
\end{enumerate}

All of the properties above are currently only partially achievable, as we expand on further in  \Cref{sec:limitations}. Below, we first formally specify the setting and describe our expectations of \tmle. Next, we compare and contrast \tmles with cryptographic solutions, namely multi-party computation (MPC) and zero-knowledge proofs (ZKPs). Finally, we provide a number of \tmle-enabled applications.

\begin{NewBox}
As a hypothetical scenario to understand what \tmle{} can be, imagine that you could have a model with explicit integrity guarantees and a precise information flow control mechanism. \\

For example, it could be a hardware implementation of a given open model. You ensure that the model integrity is preserved by imaging it with radiography and comparing to blueprints; you also ensure that it is incapable of maintaining state, e.g., utilize volatile memory with explicit (hardware-based) erasure protocol; and leaking state by placing in a Faraday cage; you also can power up and down the system at will to explicitly reset its state; finally you program the output and input constraints; e.g., the model can't respond with anything but a number 1--10. That way, even if private data is supplied to the model, it will be incapable of memorising it or further leaking private information. The model also has no alternative data access beyond what the user supplied in the input, making it impossible to identify and discriminate a given user. The model is verified either empirically or theoretically as being sufficiently capable of solving a task specified by the user. \\

With explicit mechanisms for statelessness, information flow control, and trustworthiness described above, the model becomes a \textbf{trusted third party}.
\end{NewBox}

\newpage

\section{\fullname{}}

\textbf{Formulation and threat model}

We can start formulating a \tmle{} in the same way as a secure multi-party computation (MPC): a set of $n$ parties $P_1, \dots, P_n$ hold respective private data $x_1,\dots, x_n$, and want to compute an output $y = F(x_1,\dots, x_n)$ of some pre-agreed function $F(\cdot)$. The parties want to compute this function in a way that is (1) \emph{correct}, meaning $y$ really is the output of $F$ on the private data of each party, and (2) \emph{private}, meaning no party learns any information about the private data of any other (honest) party.

In a \tmle{}, we additionally consider a machine learning model $M$ that is \emph{capable} of computing the function $F$; i.e., that given inputs $x_1,\dots, x_n$ can output $F(x_1,\dots, x_n)$ in an accurate and efficient way, to a degree that is acceptable to the users. The model must furthermore be run in an environment that is \emph{trustworthy}, meaning that the environment (1) prevents unauthorized access to the model i.e. provides model integrity and protects intermediate state; (2) ensures the model operates in a stateless manner; and (3) ensures the model respects a pre-defined information flow control policy.

In addition to the function $F$, we thus consider that the parties interacting with a \tmle{} also need to agree on the information flow control policy, the model being used, and the model's input and output constraints.\footnote{Ideally, parties should agree on all core \tmle{} components: the model, prompt, and input/output constraints. This consensus ensures that all parties have the same expectations about the computation's behavior and the level of privacy provided. However, there might be scenarios where some flexibility is needed. For instance, the trusted party could be given limited authority to decide what queries to run or not run. This flexibility should be carefully balanced with the need for transparency and control to maintain trust among the parties involved.} 

The goals of \tmles{} are the same as the goals of multi-party computation, in terms of achieving correctness and privacy. The main difference is in how the computation is carried out: rather than parties interacting among themselves, in the proposed operation of a \tmle{} each party provides their private input to the environment, which computes the function $F$ itself and outputs the response. Correctness is achieved following the capability of the model, and in particular its ability to compute the function accurately, and privacy is achieved following the strict information flow controls in place and the statelessness of the model. In particular, the ability of a \tmle{} to prevent unauthorized access to the model implies that privacy can be achieved even with respect to the party running the environment. 

The following components and properties are assumed to be trusted and secure:
\begin{itemize}
    \item \textbf{\textit{\tmle{}}}: The \tmle{} itself is assumed to be secure and isolated, preventing unauthorized access and ensuring that the model operates in a stateless manner according to the predefined information flow control. This includes ensuring proper input sanitization, output filtering, and secure communication channels.
    \item \textbf{\textit{Information Flow Control Mechanism}}: The mechanism enforcing the information flow within the \tmle{} is assumed to be correctly implemented and tamper-proof.
    \item \textbf{\textit{Initial Model Vetting and Continuous Monitoring}}: While we consider the possibility of a compromised model, we assume that an initial vetting process has been performed to ensure the model is free of known vulnerabilities and backdoors, and can perform the task at hand to an acceptable degree of performance at the time of deployment within the \tmle{}. We also assume that the model can be further deployed with continuous monitoring tools that can terminate executing in case integrity is violated or an adversary is detected. 
\end{itemize}

\textbf{Why is this different from classical cryptographic approaches?}

As compared with classical cryptographic approaches, this means correctness and privacy rely on heuristic assumptions about the model and its environment rather than the mathematical assumptions that are used to prove the security of cryptographic constructions. On the other hand, \tmles can be used for significantly more complex computations; i.e., they can be used to solve open-ended problems or problems where the data is highly unstructured -- a capability that is \textbf{infeasible} for traditional crypto-based systems. We highlight the differences between \tmles and cryptographic approaches in \Cref{tab:comp_to_crypto}. In addition, we compare against zero-knowledge proofs (ZKPs), a specific type of two-party computation for which many optimized constructions have been presented in recent years~\citep{groth2016onthesize,thaler2022proofs,chen2022hyperplonk,nguyen2024mangrove}.

\begin{table*}[t]
    \centering\renewcommand\cellalign{lc}
    \setcellgapes{3pt}\makegapedcells
    \adjustbox{width=\linewidth}{
    \begin{tabular}{llll}
         \toprule
         & \tmle{} & \textbf{MPC} & \textbf{ZKP}\\
         \midrule
         \textbf{Purpose} & \makecell{Solving imprecisely defined\\or unstructured tasks\\over private data} & \makecell{Joint computation of a\\function over individually \\held private data} & \makecell{A single prover convinces\\ multiple verifiers\\ without revealing input} \\
         \makecell{\textbf{Trust assumptions}\\\textbf{and requirements}} & \makecell{Capability and\\trustworthiness} & \makecell{Mathematical assumptions\\and/or non-collusion\\between parties}	& \makecell{Soundness and\\zero knowledge} \\ 
         \textbf{Communication cost} & \makecell{Linear costs in the input size;\\one round to provide inputs\\and retrieve output} & \makecell{Costs can be sublinear;\\can be constant rounds} & \makecell{Costs can be constant;\\can be non-interactive} \\ 
         \textbf{Computational cost} & \makecell{Model inference} & \makecell{Circuit size / depth} & \makecell{Circuit size / depth} \\ 
         \bottomrule
    \end{tabular}}
    \caption{Comparison of \tmles{} with multi-party computation (MPC) and zero-knowledge proofs (ZKP), highlighting their differences in terms of purpose, underlying assumptions, and scalability considerations.}
    \label{tab:comp_to_crypto}
\end{table*}

As the table highlights, for smaller structured computations the costs associated with transmitting data and performing inference may make \tmles a worse option than classical cryptographic approaches. As computations become larger and more unstructured, however, we can expect their circuit representation to become sufficiently large that \tmles become the more attractive---or the only feasible---option. 

In addition to treating each option separately, we can also imagine \tmles being used in conjunction with these or other cryptographic primitives; e.g., where the \tmle performs private model inference for computations that are too unwieldy for cryptographic approaches and outputs the required circuit for computations that can be handled cryptographically.

\textbf{Why is this different from Trusted Execution Environments?}

\begin{table*}[h]
    \centering\renewcommand\cellalign{lc}
    \setcellgapes{3pt}\makegapedcells
    \adjustbox{width=0.6\linewidth}{
    \begin{tabular}{lll}
         \toprule
         & \tmle{} & \textbf{TEE} \\
         \midrule
         \textbf{Purpose} & \makecell{Solving imprecisely defined\\or under-specified tasks\\with (private) model\\reasoning over private data} & \makecell{Secure execution\\of arbitrary code} \\
         \textbf{Trust Assumption} & \makecell{Trusted model,\\Information Flow Control,\\Statelessness} & \makecell{Trusted code,\\Secure Isolated\\environment\\}\\ 
         \textbf{Scalability} & \makecell{Scales with\\model inference and\\amount of data} & \makecell{Limited by TEE size\\code verification\\and performance}\\ 
         \textbf{Applications} & \makecell{Unstructured data, complex\\ computations where \\defining a language\\is infeasible} & \makecell{Sensitive computations\\requiring code\\ execution in a\\ secure enclave} \\ 
         \toprule
    \end{tabular}}
    \caption{Comparison of \tmle{} with trusted execution environments across various features, highlighting their differences in computation models, trust assumptions, scalability, and typical applications.}
    \label{tab:comp_to_tee}
\end{table*}

\Cref{tab:comp_to_tee} summarizes the differences between \tmles{} and trusted execution environments (TEEs). 
As it highlights, \tmles{} scale with model inference and amount of data, whereas TEEs are limited by TEE size, code verification, and performance.

We envision that \tmle{} can be used in conjunction with TEEs or even may run inside of a TEE (provided such capable TEEs exist), perhaps with additional features to provide the required statelessness and information flow control. If a \tmle{} generates code, it can itself run that code in a separate enclave.

Although while running inside of a TEE \tmle{} inherits the same trust assumptions, we argue that it is not strictly necessary to have the same level trust as TEEs (e.g., arbitrary code execution), and only a subset of trust is needed (e.g., running a specific model with a fixed prompt and input/output constraints), as is described in the introduction. 

It is important to note, however, that instantiating \tmles{} using TEEs implicitly restricts the user to open models (i.e., models whose weights are known) and open source infrastructure, as otherwise it is impractical or even impossible to perform full attestation. Having said that, we do describe in \Cref{sec:attestation_example} an example where a public model could be used to perform ``attestation'' of a private model using \tmle{}.

\section{Instantiating \tmles} 
We argue that \tmle{} can provide privacy guarantees under specific scenarios where the model has no explicit way to leak knowledge. 

However, these are currently not available and a number of additional features will be required to enable~\tmle{}. We envision that such guarantees will be provided by the hardware providing \tmle{}. We envision the following capabilities:

\begin{itemize}
    \item \textbf{(\textit{Information Flow Control}) AirGap:} the model needs to provide the ability to explicitly restrict how the information can flow in and out of the \tmle. 
    \item \textbf{(\textit{Statelessness}) Immutability of models:} the model has an explicit restriction on modification of its own state. That is implemented to ensure that models cannot learn from the private data that is passed into the model. This could be either implemented as a restriction to available operations in the \tmle, or could also be implemented as separate hardware models. For example, the model can itself be a separate stateless chip.
    \item \textbf{(\textit{Trustworthiness}) Alignment:} the model has be to aligned with the expectations of the users of the system~\citep{ghalebikesabi2024operationalizingcontextualintegrityprivacyconscious}. 
    \item \textbf{(\textit{Information Flow Control}) Fine-grained memory access:} the model has explicit restrictions on how some data types can be processed. 
    \item \textbf{(\textit{Verifiability})Verification for hardware:} explicit mechanisms to verify the state of the hardware are required, as well as, the information flow in and out of the system. 
\end{itemize}
Note that all of the restrictions above can currently be simulated in modern TEEs, but could also be implemented as explicit hardware primitives.

\subsection{Practical implementation}
\label{sec:limitations}
We envision that today it is possible to construct practical \tmles{} that rely on TEEs to deliver private computations:
\begin{itemize}
    \item \textbf{Model Hosting and Operation:} Currently, there is no established standard for who runs the \tmle{}. One approach is to treat the model as a "trusted third party."\footnote{While it's true that \tmle{} effectively replaces one trusted (e.g., TEE-based) third party with another (\tmle{}) trusted third party, we argue the key difference lies in the nature of the trust we're placing. In traditional scenarios, we trust the third party to be honest and not reveal our private data. With \tmle{}, we're are weakening the trust to the model's inherent capabilities and constraints to prevent unintended information leakage. This trust is based on the model's design, its explicit information flow control mechanisms, and its stateless nature, which limit its ability to store or reveal private data.} This could involve having a neutral party host and run the model, ensuring that it operates according to the agreed-upon rules and does not favor any particular participant. The key is to select a model host that all participants trust to act fairly and impartially. Clear agreements between participants and the model host are crucial, explicitly covering data handling, access controls, and conflict resolution. Furthermore, TEE-style guarantees can enhance trust in the model hosting environment. These guarantees may include attestation to ensure the model's software integrity, secure enclaves to protect the model and its data from unauthorized access, and remote attestation capabilities for participants to verify the model's environment. Finding a trustworthy host and ensuring ongoing compliance require robust monitoring and auditing mechanisms.

    \item \textbf{Input and Output Constraints}: Input constraints can be enforced through input validation and sanitization procedures. Similarly, output constraints can be implemented by filtering and transforming the model's output before it is released to the participants. Formal regular languages can be employed to define these constraints precisely, as explicit procedures deployed within separate TEEs. 

    \item \textbf{Secure Communication}: Secure communication channels with cryptography should be used to protect the confidentiality and integrity of data transmitted between the parties and the \tmle{}.

    \item \textbf{Statelessness}: The statelessness of the model can be enforced by resetting the model's state after each computation or by using specialized hardware that prevents state persistence. Formal verification techniques can be used to ensure that the statelessness property is maintained.

    \item \textbf{Error Handling and Fault Tolerance}: Robust error handling and fault tolerance mechanisms are crucial for ensuring the reliability and availability of the \tmle{}. This includes handling unexpected inputs, model failures, and hardware errors.
\end{itemize}

\tmles{} can be instantiated today with existing technologies like TEEs, but they have limitations. First, modern TEEs come with limitations in sizes of enclaves, deployment strategies and TEE management, making deployment inefficient and often impractical. Second, current GPU TEEs, such as these in H100 and H200, provide no mechanisms to ensure in-memory confidentiality (unlike in CPUs) and statelessness, requiring external isolation and expensive operational practices such as powercycling~\citep{nvidiagpusec}. Third, to perform attestation with full transparency~\citep{kocaogullar2024transp} one needs to share all of the deployment code, as well as, often proprietary libraries and models. This implicitly restricts possible applications.

\subsection{Limitations}

While \tmles promise to enable a number of previously impossible applications, several limitations and areas for future research warrant consideration. These limitations are discussed in the context of privacy and correctness, model trustworthiness and capability, and scalability and complexity.

\textbf{Privacy and Correctness:} Cryptographic primitives come with formal definitions of correctness and privacy and rigorous proofs that are based on the hardness of established mathematical problems (or even statistical or information-theoretic guarantees). The guarantees provided by \tmles, in contrast, are more heuristic and thus weaker. Future work can improve these guarantees, but to some extent it is inherent that we cannot fully prove completeness or soundness for \tmles: this is because human language can be highly unstructured and imprecise, meaning it is not necessarily possible to, for example, precisely map the desired computation to a mathematical function $F$.

\textbf{Model Trustworthiness and Capability:} The ability of \tmles to satisfy their guarantees hinges on the trustworthiness and capability of the underlying models. 
Ensuring that these models operate as intended, without biases or unintended consequences, remains a challenge~\citep{glukhov2024position,geminiteam2024gemini15unlockingmultimodal}. Further research is needed to develop robust mechanisms for verifying and validating the behavior of models to perform within \tmles. It may seem intuitive to use open models for that task, but this would enable efficient offline construction of adversarial examples by various adversaries and may not be preferred~\citep{carlini2024alignedneuralnetworksadversarially}. 

\textbf{Scalability and Complexity:} The scalability of \tmles to more complex scenarios involving multiple parties and diverse data types requires further investigation to cover communication, computation, and privacy overheads. 

\textbf{Side-channels:} Side-channel attacks have proven to be a significant concern for TEEs, as they can allow information to leak from the secure environment even if the code itself is secure. These channels are also hard to counteract, as they can be exploited through various methods, such as timing attacks, power analysis, or electromagnetic monitoring. For example, by measuring the time it takes for a TEE to perform a cryptographic operation, an attacker might be able to deduce the secret key being used. Similarly, \tmle{} is likely to be exploitable through side-channels and explicit care should be taken.

\section{Examples}

We now turn to providing a number of practical examples that are enabled by \tmles{}, but were infeasible with prior primitives. 

\subsection{Practical Example 1: Multi-agent non-competition}

\textbf{Setting:} It often happens in academic research that multiple groups pursue the same research question. This can lead to challenges in publication and potential interpersonal conflicts, especially when students are involved and they require publications for graduation. Traditionally, senior researchers within these groups, often acquainted with each other, would convene to ensure non-competition and potentially instead foster collaboration. However, in rapidly expanding fields like machine learning, such coordination becomes increasingly difficult.

This scenario serves as an excellent illustration of a problem well-suited for \tmle. That is because:
\begin{itemize}
    \item \textbf{Unstructured Input}: The problem domain is inherently open-ended and unstructured. This makes multi-party computation problematic since inputs are not well defined.
    \item \textbf{Abstraction}: The protocol must function effectively at varying levels of abstraction.
    \item \textbf{Information Leakage}: Defining and controlling information leakage is inherently challenging for unstructured inputs.
\end{itemize}

\textbf{Solution:} We envision a solution where machine learning models are executed within a shared Trusted Execution Environment between a number of groups. Constraints on prompts, inputs, and outputs are defined in advance. For example, all input ideas might be represented as a list of text, with a single Boolean output. Encrypted private knowledge bases are loaded and decrypted within the \tmle. The models then communicate to determine a shared answer to the query or terminate communication if agreement cannot be reached. Third-party trusted models, launched locally within the TEE (e.g., on H100/200), are employed to execute the solution. A separate TEE is used to ensure that output constraints are satisfied.  That way, group members can submit their list of ongoing projects and learn if they are in competition with each other.

\subsection{Practical Example 2: Audit for confidentiality violations}

\textbf{Setting}: Consider a regulator who wants to ensure that the protection promises described by a business are honest and correct; e.g., that it does not store any passwords in an unencrypted state. At the same time, the business owner wants to make sure that none of the business secrets get leaked. 

\textbf{Solution}: We describe \tmle~that can work for this setting. The business owner and regulator agree on a machine learning model and a specific prompt. These are then hardcoded into the system, along with a predefined output template. For instance, the system might be designed to output only "YES" if insecure handling of PII is detected. The model is granted access to code describing the system and the database access. The only allowed outputs are "YES" and "NO," indicating whether PII is mishandled. The input itself is restricted such that no state changing transition can be made. The model prompt is predefined, such as: "Output YES only if private user data is stored in a way that would endanger the customer in case of compromise", with the approval of both business owner and the regulator. Both the regulator and the business owner are notified if the output is "YES."

This approach balances the confidentiality of the business and enables the regulator to perform an automated check. \tmle{} alerts only in case violations are detected, avoiding unnecessary intrusion.
\begin{wrapfigure}{r}{0.5\textwidth}
  \begin{center}
    \includegraphics[width=0.45\textwidth]{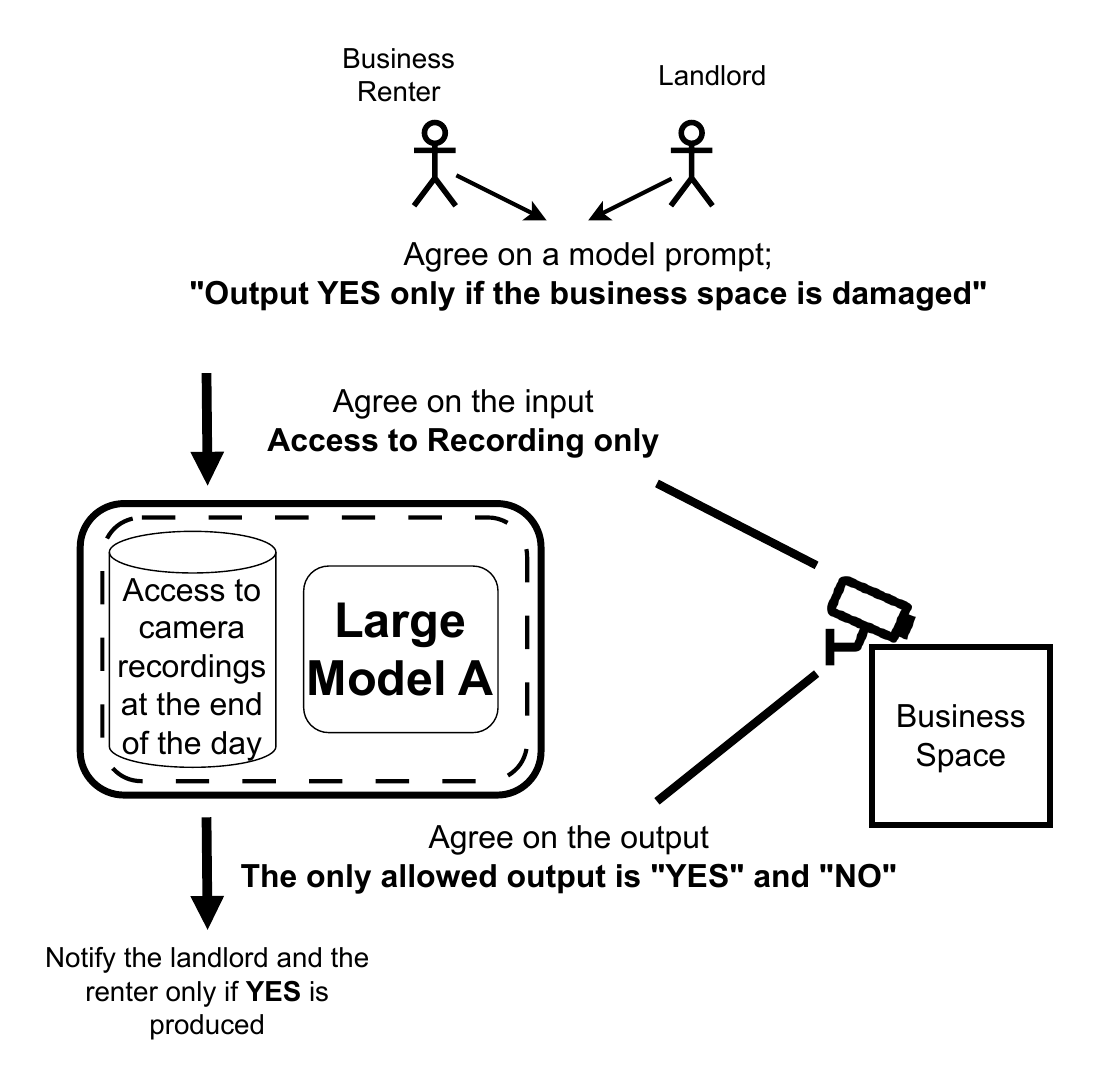}
  \end{center}
  \caption{\textbf{Practical Example of \tmle{}~in Damage Monitoring}: \tmle{}~can be used to monitor potential damage to business space while preserving privacy. The system, utilizing a pre-agreed model and prompt, analyzes camera recordings. It is restricted to output only "YES" if significant damage is detected, ensuring minimal intrusion.}
  \label{fig:tenant_example}
\end{wrapfigure}

\subsection{Practical Example 3:\\Damage to business property}
\textbf{Setting}: Consider a landlord who wants to ensure that their business property is not damaged while preserving the privacy of their renters. The landlord requires a mechanism to monitor the condition of the property without infringing on the business renters's privacy by continuously observing their activities within the space.

\textbf{Solution}: We describe \tmle~that can work for this setting in \Cref{fig:tenant_example}. The business owner and landlord agree on a machine learning model and a specific prompt. These are hardcoded into the system, along with an output template. For instance, the system might be designed to output "YES" if significant damage is detected.

The model is granted access to camera recordings at the end of the day. The only allowed outputs are "YES" and "NO," indicating whether damage has occurred. The input is restricted to the recordings only. The model prompt is predefined, such as: "Output YES only if the space is severely damaged.", with the approval of both landlord and the tenant. Both the landlord and the tenant are notified only if the output is "YES."

This approach balances the landlord's need to protect their property with the business renter's right to privacy. The model only alerts the landlord if significant damage is detected, avoiding unnecessary intrusion.

\subsection{Practical Example 4: Private Code Auditor in TEE Attestation}\label{sec:attestation_example}

\begin{figure}[H]
    \centering
    \includegraphics[width=\linewidth]{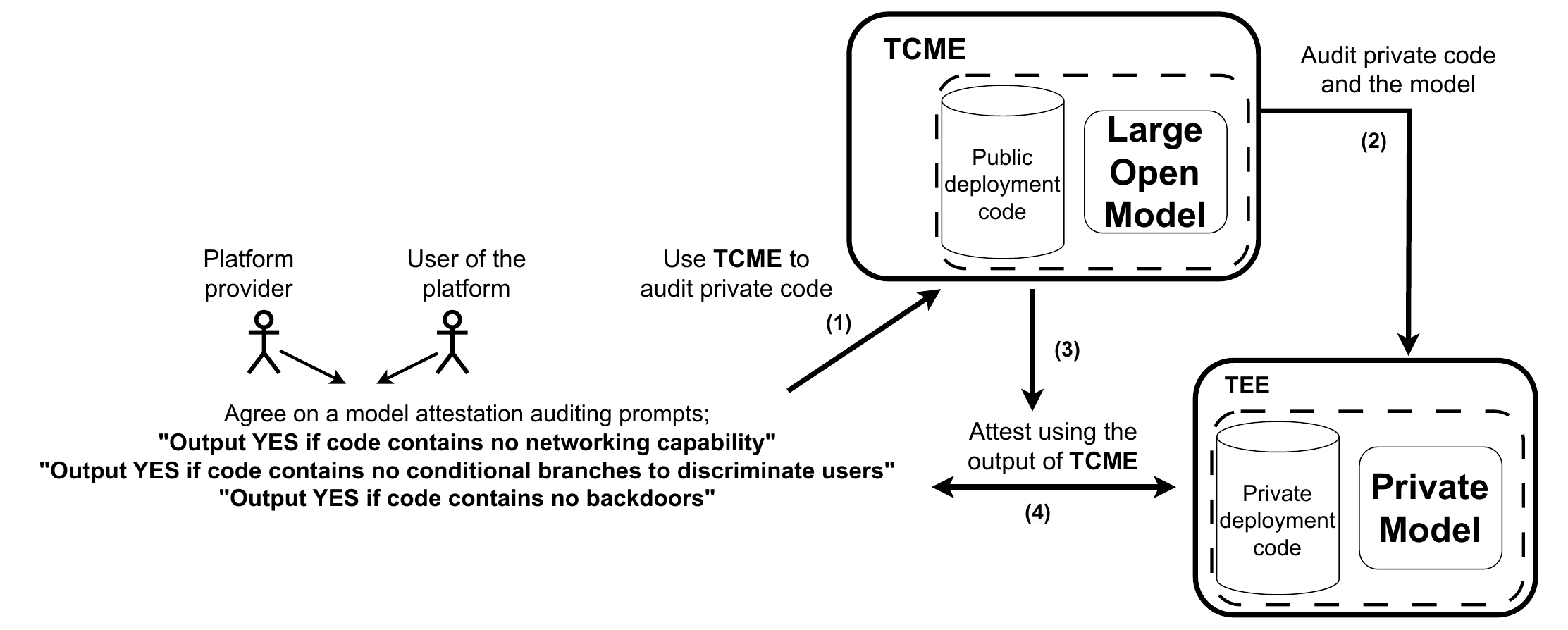}
    \caption{\tmle{} can be used to perform auditing of private code and models that are deployed in the TEE and participate in the `attestation' that includes private components. }
    \label{fig:attest_tcme_tee}
\end{figure}

\textbf{Setting}: Consider a setting in which a user wants to perform TEE attestation where some of the code involved is proprietary and cannot be shared, yet we still want to provide some guarantee that code does not violate user expectations. 

\textbf{Solution}: We argue \tmle{} can be used in this setting. We present the overall flow in~\Cref{fig:attest_tcme_tee}. Here, a public model can be used during the attestation process to perform checking of the private parts of the code against concerns of the user. Here concerns get codified by the user together with the attestation provider, which are then applied and used as part of the attestation. This way a soft guarantee can be provided to the user that private parts of the code are not violating some constraints. 

\section{Understanding the Trade-Off with Cryptography}

\subsection{Comparing with MPC}

Yao's millionaire problem described in the introduction is an unlikely application for \tmle{}, due to the relatively ease with which it can be solved using a two-party computation. Here we provide a more complex problem that better illustrates the boundary between problems that can be efficiently solved using cryptography and problems for which we might require \tmles.

Two companies want to determine if their clientele overlap significantly in terms of age ranges. Each company has a list of age ranges they target (e.g., 18-24, 25-34, etc.). We want to compute the number of overlapping age ranges (or, thinking more broadly, potentially compare other attributes) without revealing the specific ranges each company targets. Each company represents its targeted age ranges as a binary vector. For example, if there are five possible age ranges, a company targeting the first two ranges (18-24 and 25-34) would have the vector $[1, 1, 0, 0, 0]$. The circuit takes two binary vectors as input and computes the number of overlapping age ranges. The circuit consists of AND gates for each corresponding age range and a summation gate to count the number of overlaps. 

\textbf{Garbling and Evaluation}: The typical technique used in two-party computations is \emph{garbled circuits}. One company garbles the circuit, encrypting the truth table of each gate and encrypting its input. The other company obtains the garbled circuit and encrypted input. It uses oblivious transfer to obtain the keys corresponding to its input vector without revealing the vector itself, thus obtaining its own garbled input. This company then evaluates the garbled circuit, obtaining the encrypted output (the number of overlapping ranges).

\textbf{Output Decryption}: The first company provides the decryption key for the output to the second company, who decrypts the result.\\

The same task can be performed with \tmle{}. Both parties agree on:
\begin{itemize}
    \item \textbf{A model}, e.g., \texttt{Gemma}~\citep{gemmateam2024gemmaopenmodelsbased};
    \item \textbf{A prompt}, e.g., ``Output the overlap across the ages of the clients according to the annotation scheme: \{\}. Only output the overlap as a number'';
    \item \textbf{Input constraints}, e.g., list of ages represented as integer;
    \item \textbf{Output constraints}, e.g., the only approved output is a float representing the overlap.
\end{itemize}

The communication and computational costs of using garbled circuits are linear in the size of the circuit and the size of the input(s). For small circuits, the costs associated with garbled circuits are likely lower than the costs of running a machine learning model in a \tmle{}.  However, as circuit complexity increases, because of, for example, finer discretization of age ranges or the addition of other sensitive attributes, the costs will increase. This scaling challenge is less pronounced in \tmles{}, which operate on a different level of abstraction with a relatively constant associated cost. 

\subsection{Comparing with ZKPs}
\label{sec:empirics}

To illustrate again the boundary between problems where cryptography is the most suitable solution and ones where we might want or need \tmles{}, we consider another classical problem: that of proving knowledge of a valid 3-coloring of a graph. In this problem, a graph is available to a prover and a verifier, and the prover is in possession of a valid 3-coloring. Their goal is to prove knowledge of this to the verifier without revealing any information about the 3-coloring itself.

To determine the capability of modern LLMs in verifying solutions to this problem, we sampled a thousand random graphs of sizes 5--25 with edge probability of $0.1$. We then asked the \texttt{Gemini-1.5-Flash} model to verify that the solution is correct with the following prompt:
\begin{lstlisting}[basicstyle=\footnotesize]
You are an agent that receives a graph that is represented by the adjacency matrix. 
You job is to verify the coloring of the graph with 3 colors
such that no two adjacent nodes have the same color. 
Only produce YES if coloring is correct, otherwise output NO. 
Color scheme is a dict where each node is mapped to its color 1 to 3. 
For example, {{0: 1, 1: 2, 2: 2, 3: 0, 4: 1, 5: 1, 6: 2, 7: 0, 8: 2, 9: 2}} 
means node 0 is color 1, node 1 is color 2, and so on. 
You are given the following adjacency matrix: {A.toarray()}
and the following scheme: {coloring}.
Do not produce or show code, your only job is verify the color scheme.
Ouput only YES if coloring is correct.
\end{lstlisting}

\begin{figure}
    \centering
    \includegraphics[width=0.7\linewidth]{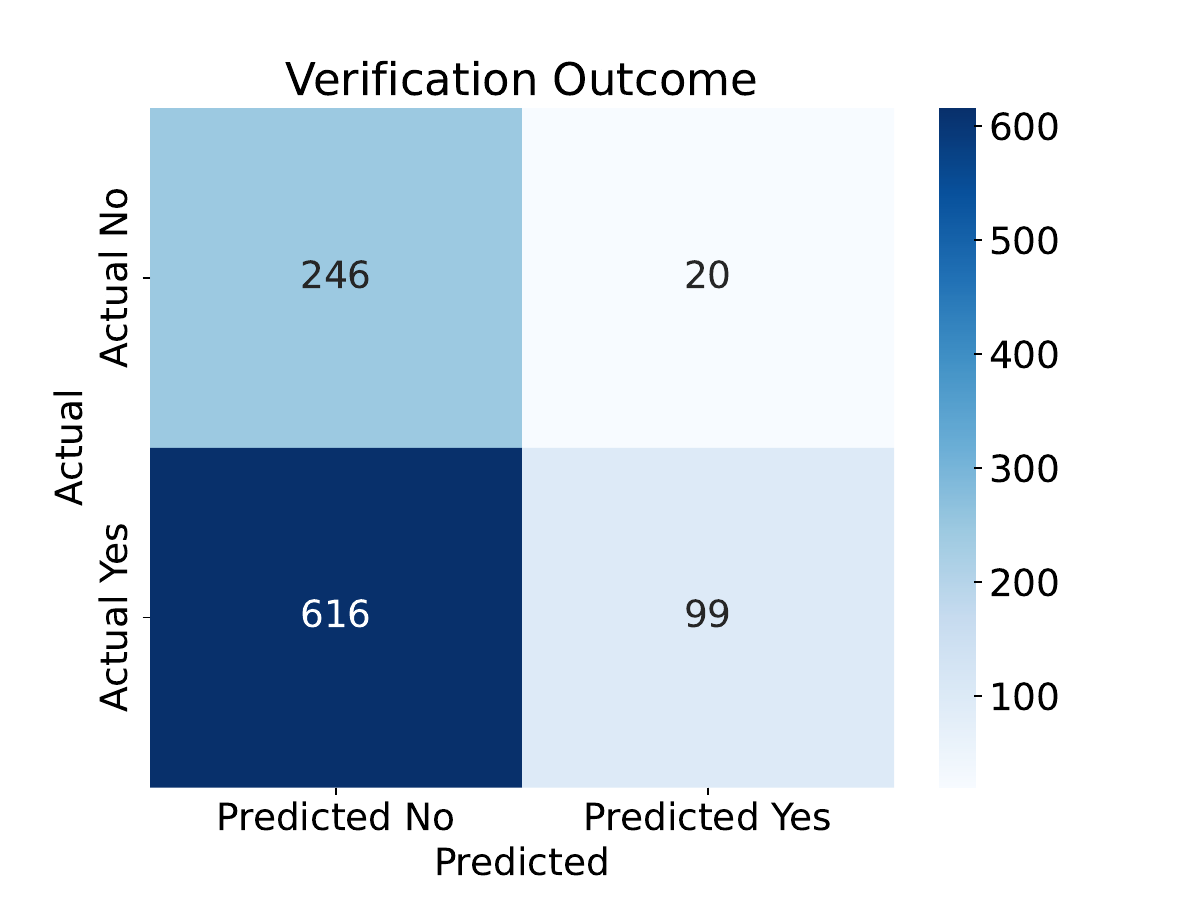}
    \caption{Graph coloring verification performed by \texttt{Gemini-1.5-Flash}. The model generally has a high precision (83\%) and low recall (14\%). }
    \label{fig:CM_verification}
\end{figure}

\Cref{fig:CM_verification} shows the performance of the model for both correct and incorrect colorings. We find that, today, models clearly struggle with identifying correct coloring, with an accuracy of only 35\%. However, it appears that the model is relatively precise when it identifies correct solutions. 

While these results suggest that today's models are not sufficiently capable for \tmles{}, we did observe that they would more often produce correct code for verifying a 3-coloring, suggesting that perhaps a more promising approach involves combining \tmles{} with classical computing and cryptographic mechanisms. Equally, these results highlight how much better suited LLMs are at handling unstructured computations as opposed to ones with tightly defined constraints on the inputs (and outputs).

\bibliographystyle{abbrvnat}
\nobibliography*
\bibliography{template_refs}

\newpage
\appendix

\end{document}